\documentclass[sigconf,screen]{acmart}

\AtBeginDocument{%
  \providecommand\BibTeX{{%
    \normalfont B\kern-0.5em{\scshape i\kern-0.25em b}\kern-0.8em\TeX}}}

\setcopyright{none}
\setcopyright{acmcopyright}
\acmPrice{15.00}
\acmDOI{}
\acmYear{2020}
\copyrightyear{2020}
\acmISBN{}
\acmConference[ASE 2020]{Proceedings of the 35th IEEE/ACM International Conference on Automated Software Engineering}{September 21--25, 2020}{Melbourne, Australia}
\acmBooktitle{Proceedings of the 35th IEEE/ACM International Conference on Automated Software Engineering (ASE 2020), September 21--25, 2020, Melbourne, Australia}

\usepackage{calligra}
\usepackage{algorithm}
\usepackage{algorithmicx}
\usepackage{algpseudocode}

\usepackage{booktabs}
\usepackage{moreverb}
\usepackage{fontenc}
\usepackage{amsmath}
\usepackage{amssymb}
\usepackage{fancybox}
\usepackage{color}
\usepackage{colortbl}
\usepackage{array}
\usepackage{multirow}
\usepackage{multicol}
\usepackage{listings}
\usepackage{makecell}
\usepackage{graphicx}
\usepackage{setspace}
\usepackage[multiple]{footmisc}
\usepackage{wasysym}

\usepackage{threeparttable}

\usepackage{balance}
\usepackage{csvsimple}
\usepackage{longtable}
\usepackage[skip=0pt,labelfont=bf]{caption}
\usepackage{url}
\usepackage{lscape}
\usepackage{rotating}
\usepackage{tikz}
\usepackage[normalem]{ulem}
\usepackage{enumitem}
\usepackage{subfigure}

\usepackage[most]{tcolorbox}

\usepackage{threeparttable}

\usepackage{marvosym}
\usepackage{pifont}
\usepackage{courier}

\algnewcommand\algorithmicforeach{\textbf{for each}}
\algdef{S}[FOR]{ForEach}[1]{\algorithmicforeach\ #1\ \algorithmicdo}

\newcolumntype{L}[1]{>{\raggedright\let\newline\\\arraybackslash\hspace{0pt}}m{#1}}
\newcolumntype{C}[1]{>{\centering\let\newline\\\arraybackslash\hspace{0pt}}m{#1}}
\newcolumntype{R}[1]{>{\raggedleft\let\newline\\\arraybackslash\hspace{0pt}}m{#1}}

\definecolor{codegreen}{rgb}{0,0.6,0}
\definecolor{codered}{rgb}{1,0,0}
\definecolor{codegray}{rgb}{0.5,0.5,0.5}
\definecolor{codepurple}{rgb}{0.58,0,0.82}
\definecolor{backcolour}{rgb}{0.95,0.95,0.92}
\definecolor{lightgray}{gray}{0.9}

\lstdefinestyle{mystyle}{
    commentstyle=\color{codegreen},
    keywordstyle=\color{magenta},
    numberstyle=\small\color{black},
    stringstyle=\color{codepurple},
    basicstyle=\scriptsize\ttfamily,
    breakatwhitespace=false,
    breaklines=true,
    captionpos=b,
    keepspaces=true,
    showspaces=false,
    showstringspaces=false,
    showtabs=false,
    tabsize=2
}

\lstdefinelanguage{diff}{
  morecomment=[f][\color{blue}]{@@},     
  morecomment=[f][\color{red}]-,         
  morecomment=[f][\color{codegreen}]+,       
  morecomment=[f][\color{red}]{---}, 
  morecomment=[f][\color{codegreen}]{+++},
}

\setlist{noitemsep} 

\definecolor{darkpastelred}{rgb}{0.76, 0.23, 0.13}
\definecolor{ao(english)}{rgb}{0.0, 0.5, 0.0}

\definecolor{darkpastelred}{rgb}{0.76, 0.23, 0.13}
\definecolor{ao(english)}{rgb}{0.0, 0.5, 0.0}

\definecolor{yellow}{RGB}{255,255,153}
\definecolor{grey}{RGB}{224,224,224}

\newboolean{showcomments}
\setboolean{showcomments}{true}
\ifthenelse{\boolean{showcomments}}
 { \newcommand{\mynote}[2]{
      \fbox{\bfseries\sffamily\scriptsize#1}
        {\small$\blacktriangleright$\textsf{\emph{#2}}$\blacktriangleleft$}}}
        { \newcommand{\mynote}[2]{}}

\definecolor{DarkOrange}{rgb}{0.8,0.3,0.0}
\definecolor{DarkCyan}{rgb}{0.0, 0.55, 0.55}
\definecolor{DarkCyel}{rgb}{1.0, 0.49, 0.0}
\definecolor{yellow-green}{rgb}{0.6, 0.8, 0.2}

\newcolumntype{?}{!{\vrule width 1pt}}

\newcommand{\find}[1]{
\begin{tcolorbox}[leftrule=0.5mm,toprule=0mm,bottomrule=0mm,left=0.7pt,right=0.7pt,top=0.2pt,bottom=0.2pt]
\em #1
\end{tcolorbox}
}

\begin{document}
\title{Evaluating Representation Learning of Code Changes for Predicting Patch Correctness in Program Repair}

\author{Haoye Tian}
\email{haoye.tian@uni.lu}
\affiliation{%
   \institution{University of Luxembourg}
 	\country{Luxembourg}
}
\author{Kui Liu}\authornote{Corresponding author.}
\email{kui.liu@nuaa.edu.cn}
\affiliation{%
   \institution{Nanjing University of Aeronautics and Astronautics}
 	\country{China}
}

\author{Abdoul Kader Kabor{\'e}}
\author{Anil Koyuncu}
\email{{abdoulkader.kabore, anil.koyuncu}@uni.lu}
\affiliation{%
   \institution{University of Luxembourg}
 	\country{Luxembourg}
}

\author{Li Li}
\email{li.li@monash.edu}
\affiliation{%
   \institution{Monash University}
   \country{Australia}
}

\author{Jacques Klein}
\email{jacques.klein@uni.lu}
\affiliation{%
   \institution{University of Luxembourg}
 	\country{Luxembourg}
}

\author{Tegawend{\'e} F. Bissyand{\'e}}
\email{tegawende.bissyande@uni.lu}
\affiliation{%
   \institution{University of Luxembourg}
 	\country{Luxembourg}
}

\begin{abstract}
A large body of the literature of automated program repair develops approaches where patches are generated to be validated against an oracle (e.g., a test suite). 
Because such an oracle can be imperfect, the generated patches, although validated by the oracle, may actually be incorrect. 
While the state of the art explore research directions that require dynamic information or rely on manually-crafted heuristics, we study the benefit of learning code representations to learn deep features that may encode the properties of patch correctness. 
Our work mainly investigates different representation learning approaches for code changes to derive embeddings that are amenable to similarity computations. 
We report on findings based on embeddings produced by pre-trained and re-trained neural networks. 
Experimental results demonstrate the potential of embeddings to empower learning algorithms in reasoning about patch correctness: a machine learning predictor with BERT transformer-based embeddings associated with logistic regression yielded an AUC value of about 0.8 in predicting patch correctness on a deduplicated dataset of 1000 labeled patches. Our study shows that learned representations can lead to reasonable performance when comparing against the state-of-the-art, PATCH-SIM, which relies on dynamic information. These representations may further be complementary to features that were carefully (manually) engineered in the literature. 
\vspace{-5mm}
\end{abstract}

\settopmatter{printacmref=true}

%



\maketitle


\section{Introduction}
\label{sec:intro}

Automation in software engineering has recently reached new heights with the promising results recorded in the research direction of automated program repair (APR)~\cite{le2019automated,monperrus2018automatic}. 
While a few techniques try to model program semantics and synthesize execution constraints towards producing quality patches, they often fail to scale to large  programs. 
Instead, the large majority of research contributions~\cite{monperrus2018living} explore search-based approaches where patch candidates are generated and then validated against an oracle.

In the absence of strong program specifications, test suites represent affordable approximations that are widely used as the oracle in APR. 
In their seminal approach to test-based APR, Weimer~{\em et~al.}~\cite{weimer2009automatically} considered that a patch is acceptable if it makes the program pass all test cases in the test suite. 
Since then, a number of studies~\cite{qi2015analysis,smith2015cure} have explored the {\em overfitting} problem in patch validation: a given patch is synthesized to pass a test suite and yet is incorrect with respect to the intended program specification. 
Since limited test suites only weakly approximate program specifications, a patched program can indeed satisfy the requirements encoded in the test cases, and present a behavior outside of those tests that are significantly different from the  behavior initially expected by the developer.

Overfitting patches constitute a key challenge in generate-and-validate APR approaches. Recent evaluation campaigns~\cite{liu2019avatar,liu2019tbar,jiang2018shaping,saha2019harnessing,wen2018context,liu2018lsrepair,koyuncu2019ifixr,liu2019you,koyuncu2020fixminer} on APR systems are stressing on the importance of estimating the correctness ratio among the valid patches that can be found. To improve this ratio, researchers are investigating several research directions. We categorize them in three main axes that focus on actions before, during or after patch generation:
\begin{itemize}[leftmargin=*]
	\item {\em test-suite augmentation:} Yang~{\em et~al.}~\cite{yang2017better} proposed to generate better test cases to enhance the validation of patches, while Xin and Reiss~\cite{xin2017identifying} opted for increasing test inputs. 
	\item {\em post-processing of generated patches:} Long and Rinard~\cite{long2016automatic} studied some heuristics to discard patches that are likely overfitting.
	\item {\em curation of repair operators:} approaches such as CapGen~\cite{wen2018context} successfully demonstrated that carefully-designed (e.g., fine-grained fix ingredients) repair operators can lead to more correct patches.
\end{itemize}

\vspace{-0.5mm}
Our work is related to the latter thrust. 
So far, the state-of-the-art works targeting the identification of patch correctness are mainly implemented based on computing the similarity of test case execution traces~\cite{xiong2018identifying}. Ye~{\em et~al.}~\cite{ye2019automated} followed up by presenting preliminary results suggesting that statically-extracted code features at the syntax level could be used to predict overfitting patches.
While such an approach is appealing, the feature engineering effort can be huge when researchers target generalizable approaches.
To cope with this problem, Csuvik~{\em et~al.}~\cite{csuvik2020utilizing} have proposed a preliminary small-scale study on the use of embeddings: leveraging pre-trained natural language sentence embedding models, they claim to have been able to filter out 45\% incorrect patches generated for 40 bugs from the QuixBugs dataset~\cite{ye2019comprehensive}.

{\bf This paper.} Embeddings have been successfully applied to various prediction tasks in software engineering research~\cite{soto2018using,wang2016automatically,liu2019learning,allamanis2014learning}. For patch correctness prediction, the literature does not yet provide extensive experimental results to guide future research. Our work fills this gap.   
We investigate in this paper the feasibility of leveraging advances in deep representation learning to produce embeddings that are amenable to reasoning about correctness. 
\begin{itemize}[leftmargin=*]\vspace{-0.5mm}
	\item[\ding{182}] We investigate different representation learning models adapted to natural language tokens and source code tokens that are more specialized to code changes. Our study considers both pre-trained models and the retraining of models.
	\item[\ding{183}] We empirically investigate whether, with learned representations, the hypothesis of minimal changes incurred by correct patches remains valid: experiments are performed to check the statistical difference between similarity scores yielded by correct patches and those yielded by incorrect patches. 
	\item[\ding{184}] We run exploratory experiments assessing the possibility to select cutoff similarity scores between learned representations of buggy code and patched code fragments for heuristically filtering out incorrect patches.
	\item[\ding{185}] Finally, we investigate the discriminative power of deep learned features in a classification training pipeline aimed at learning to predict patch correctness.
\end{itemize}

\section{Background}
\label{sec:bg}
Our work deals with various concepts and techniques from the fields of program repair and machine learning. We present the relevant details in this section to facilitate readers' understanding of our study design and the scope of our experiments.

\vspace{-2mm}
\subsection{Patch Plausibility and Correctness}
Defining patch correctness is a non-trivial challenge in APR. 
Until the release of empirical investigations by Smith~{\em et~al.}~\cite{smith2015cure}, actual correctness (w.r.t. program behavior intended by developers) was seldom used as a performance criterion of patch generation systems. 
Instead, experimental results were focused on the number of patches that make the program pass all test cases. Such patches are actually only {\bf plausible}. 
Qi~{\em et~al.}~\cite{qi2015analysis} demonstrated in their study that an overwhelming majority of plausible patches generated by GenProg~\cite{le2012genprog}, RSRepair~\cite{qi2014strength} and AE~\cite{weimer2013leveraging}) are overfitting the test suite while actually being incorrect.
To improve the probability of program repair systems to generate {\bf correct} patches, researchers have mainly invested in strengthening the validation oracle (i.e., the test suites).
Opad~\cite{yang2017better}, DiffTGen~\cite{xin2017identifying}, UnsatGuided~\cite{yu2019alleviating}, PATCH-SIM/TEST-SIM~\cite{xiong2018identifying} generate new test inputs that trigger behavior cases which are not addressed by APR-generated patches.

More recent works~\cite{ye2019automated,csuvik2020utilizing} are starting to investigate static features and heuristics (or machine learning) to build predictive models of patch correctness. Ye~{\em et~al.}~\cite{ye2019automated} presented the ODS approach which relates to our study since it investigated machine learning with static features extracted from Java program patches. Their approach however builds on carefully hand-crafted features, which may not generalize to other programming languages or even to varied datasets.  The study of Csuvik~{\em et~al.}~\cite{csuvik2020utilizing} is also close to ours as it explores BERT embeddings to define similarity thresholds. Their work remains preliminary (it does not investigate the discriminative power of features) and has been performed at a small scale (single pre-trained model on 40 one-line bugs from simple programs).

\vspace{-2mm}
\subsection{Distributed Representation Learning}
Learning distributed representations have been widely used to advance several machine learning tasks. In particular, in the field of Natural Language Processing embedding techniques such as  Word2Vec~\cite{le2014distributed}, {\bf Doc2Vec}~\cite{le2014distributed} and {\bf BERT}~\cite{devlin2019bert} have been successfully applied for different semantics-related tasks. 
By building on the hypothesis of code naturalness~\cite{hindle2012naturalness,allamanis2018survey}, a number of software engineering research works have also leveraged the aforementioned approaches for learning distributed representations of code. Alon~{\em et~al.}~\cite{alon2019code2vec} have then proposed {\bf code2vec}, an embedding technique that explores AST paths to take into account structural information in code.  More recently, Hoang~{\em et~al.}~\cite{hoang2020cc2vec} have proposed {\bf CC2Vec}, which further specializes to code changes.

Our work explores different techniques across the spectrum of distributed representation learning. We therefore consider four variants from the seemingly-least specialized to code (i.e., Doc2Vec) to the state of the art for code change representation (i.e., CC2Vec).

\vspace{-1mm}
\subsubsection{\bf Doc2Vec}
Doc2Vec~\cite{le2014distributed} is an unsupervised framework mostly used to learn continuous distributed vector representations of sentences, paragraphs and documents, regardless of their lengths.
It works on the intuition, inspired by the method of learning word vectors~\cite{mikolov2013efficient}, that the document representation should be good enough to predict the words in the document
Doc2Vec has been applied in various software engineering tasks.
For example, 
Wei and Li~\cite{wei2017supervised} leveraged Doc2Vec to exploit deep lexical and syntactical features for software functional clone detection.
Ndichu~{\em et~al.}~\cite{ndichu2019machine} employed Doc2Vec to learn code structure representation at AST level to predict JavaScript-based attacks.

\vspace{-1mm}
\subsubsection{\bf BERT}
BERT~\cite{devlin2019bert} is a language representation model that has been introduced by an AI language team in Google.
BERT is devoted to pre-train deep bidirectional representations from unlabelled texts. 
Then a pre-trained BERT model can be fine-tuned to accomplish various natural language processing tasks such as question answering or language inference.
Zhou~{\em et~al.}~\cite{zhou2019lancer} employed a BERT pre-trained model to extract deep semantic features from code name information of programs in order to perform code recommendation.
Yu~{\em et~al.}~\cite{yu2020order} even leveraged BERT on binary code to identify similar binaries.

\vspace{-1mm}
\subsubsection{\bf code2vec}
code2vec~\cite{alon2019code2vec} is an attention-based neural code embedding model developed to represent code fragments as continuous distributed vectors, by training on AST paths and code tokens. 
 Its embeddings have notably been used to predict the semantic properties of code fragments~\cite{alon2019code2vec}, in order, for instance, to predict method names. 
In a recent work, however, Kang~{\em et~al.}~\cite{kang2019assessing} reported an empirical study, which highlighted that the yielded token code2vec embeddings may not generalize to other code-related tasks  such as code comment generation, code authorship identification or code clone detection. code2vec remains however the state of the art in code embeddings: 
Compton~{\em et~al.}~\cite{compton2020embedding} recently leveraged code2vec to embed Java classes and learn code structures for the task of variable naming obfuscation.

\vspace{-1mm}
\subsubsection{\bf CC2Vec}
CC2Vec~\cite{hoang2020cc2vec} is a specialized hierarchical attention neural network model which learns vector representations of code changes (i.e., patches) 
guided by the associated commit messages (which is used as a semantic representation of the patch). 
As the authors demonstrated in their in large empirical evaluation, CC2Vec presents promising performance on commit message generation, bug fixing patch identification, and just-in-time defect prediction.


\section{Study Design}
\label{sec:exp}
First, we overview the research questions that we investigate. Then we present the datasets that are leveraged to answer these research questions. Finally, we discuss the actual training of (or use of pre-trained) models for embedding the code changes.

\vspace{-1mm}
\subsection{Research Questions}

\begin{description}
	\item {\em {\bf RQ1:} Do different representation learning models yield comparable distributions of similarity values between buggy code and patched code?} 
	A widespread hypothesis in program repair is that bug fixing generally induce minimal changes~\cite{chen2017testing,xiong2017precise,jiang2019inferring,jiang2018shaping,liu2019avatar,liu2019tbar,wen2018context,weimer2009automatically,liu2018closer,martinez2015mining,barr2014plastic}. We propose to investigate whether embeddings can be a reliable means for assessing the extent of changes through computation of cosine similarity between vector representations.
	\item {\em {\bf RQ2:} To what extent similarity distributions can be generalized for inferring a cutoff value to filter out incorrect patches?} 
	Following up on RQ1, We propose in this research question to experiment ranking patches based on cosine similarity of their vector representations, and rely on naively-defined similarity thresholds to decide on filtering of incorrect patches.
	\item {\em {\bf RQ3:} Can we learn to predict patch correctness by training classifiers with code embeddings input?} 
	We investigate whether deep learned features are indeed relevant for building machine learning predictors for patch correctness. 
\end{description}
%
%

\vspace{-2mm}
\subsection{Datasets}
We collect patch datasets by building on previous efforts in the community. An initial dataset of correct patches is collected by using five literature benchmarks, namely Bugs.jar~\cite{saha2018bugs}, Bears~\cite{madeiral2019bears}, Defects4J~\cite{just2014defects4j}, QuixBugs~\cite{lin2017quixbugs} and ManySStuBs4J~\cite{karampatsis2020how}. These are developer patches as committed in open source project repositories.

We also consider patches generated by APR tools integrated into the \texttt{RepairThemAll} framework. We use all patch samples released by Durieux~{\em et~al.}~\cite{durieux2019empirical}. This only includes sample patches that make the programs pass all test cases. They are thus plausible. However, no validation information on correctness was given. In this work, we proceed to manually validate the generated patches, among which we identified 900 correct patches. The correctness validation follows the criteria defined by Liu~{\em et~al.}~\cite{liu2020efficiency}.

In a recent study on the efficiency of program repair, Liu~{\em et~al.}~\cite{liu2020efficiency} released a labeled dataset of patches generated by 16 APR systems for the Defects4J bugs. We consider this dataset as well as the labeled dataset that was used to evaluate the PATCH-SIM~\cite{xiong2018identifying} approach. 

Overall, Table~\ref{tab:datasets} summarizes the data sets that we used for our experiments. Each experiment in Section~\ref{sec:eval} has specific requirements on the data (e.g., large patch sets for training models, labeled datasets for benchmarking classifiers, etc.). For each experiment, we will recall which sub-dataset has been leveraged and why. 

\begin{table}[h!t]
	\centering
	\caption{Datasets of Java patches used in our experiments.}	
	\resizebox{1\linewidth}{!}{
	\begin{threeparttable}
	\begin{tabular}{L{26mm}|C{15mm}C{15mm}C{12mm}r}
			\toprule
		Subjects &  contains incorrect patches & contains correct patches & labelled dataset & \# Patches \\ \hline
		Bears~\cite{madeiral2019bears}  & No & Yes & - & 251 \\
		Bugs.jar~\cite{saha2018bugs}  & No & Yes & - & 1,158 \\
		Defects4J~\cite{just2014defects4j}$^\dagger$  & No & Yes & - & 864 \\
		ManySStubBs4J~\cite{karampatsis2020how}  & No & Yes & - & 34,051 \\
		QuixBugs~\cite{lin2017quixbugs}  & No & Yes & - & 40 \\\hline\hline
		RepairThemAll~\cite{durieux2019empirical} & Yes & Yes & No$^\ddagger$ & 64,293 \\
		Liu~{\em et~al.}~\cite{liu2020efficiency} & Yes & Yes & Yes & 1,245 \\	
		Xiong~{\em et~al.}~\cite{xiong2018identifying} & Yes & Yes & Yes & 139 \\\hline\hline
		{\bf Total} & & & & 102,041 \\
			\bottomrule
	\end{tabular}
	{$^\dagger$The latest version 2.0.0 of Defects4J is considered in this study.\\
	$^\ddagger$The patches are not labeled in~\cite{durieux2019empirical}. We support the labeling effort in this study by comparing the generated patches against the developer patches. The 2,918 patches for IntroclassJava in~\cite{durieux2019empirical} are also excluded from our study since IntroClassJava is a lab-built Java benchmark transformed from the C program bugs in small student-written programming assignments from IntroClass~\cite{le2015manybugs}.}
	\end{threeparttable}
	}
\label{tab:datasets}
\end{table}

\vspace{-2mm}
\subsection{Model input pre-processing}
Samples in our datasets are patches such as the one presented in Figure~\ref{fig:chart1} extracted from the Defects4J dataset.
Our investigations with representation learning however require input data about the buggy and patched code. A straightforward approach to derive those inputs would be to consider the code files before and after the patch. Unfortunately, depending on the size of the code file, the differences could be too minimal to be captured by any similarity measurement. To that end, we propose to focus on the code fragment that appears in the patch. Thus, to represent the buggy code fragment (cf. Figure~\ref{fig:chart1-buggy}), we keep all removed lines (i.e., starting with `-') as well as the patch context lines (i.e., those not starting with either `-', `+' or `@'). Similarly, the patched code fragment (cf. Figure~\ref{fig:chart1-patched}) is represented by added lines (i.e., starting with '+') as well as the same context lines. 
Since tool support for the representation learning techniques BERT, Doc2Vec, and CC2Vec require each input sample to be on a single line, we flatten multi-line code fragments into a single line.

In contrast to BERT, Doc2Vec, and CC2Vec, which can take as input some syntax-incomplete code fragments, code2vec requires the fragment to be fully parsable in order to extract information on Abstract Syntax Tree paths. Since patch datasets include only text-based diffs, code context is generally truncated and is likely not parsable. 
However, as just explained, we opt to consider only the removed/added lines to build the buggy and patched code input data. By doing so, we substantially improved the success rate of the JavaExtractor tool used to build the tokens in the code2vec pipeline.

\begin{figure}[!t]
    \centering\vspace{2mm}
    \scriptsize
\begin{lstlisting}[language=diff,linewidth={\linewidth},frame=tb,basicstyle=\ttfamily, basicstyle=\footnotesize]
--- source/org/jfree/chart/renderer/category/AbstractCategoryItemRenderer.java
+++ source/org/jfree/chart/renderer/category/AbstractCategoryItemRenderer.java
@@ -1795,6 +1795,6 @@ public abstract class AbstractCategoryItemRenderer
         int index = this.plot.getIndexOf(this);
         CategoryDataset dataset = this.plot.getDataset(index);
-        if (dataset != null) {
+        if (dataset == null) {
             return result;
         }
\end{lstlisting}
    \caption{Example of a patch for the Defects4J bug Chart-1.}
    \label{fig:chart1}
\end{figure}

\vspace{-1mm}
\begin{figure}[!t]
    \centering
    \scriptsize
\begin{lstlisting}[language=Java,linewidth={\linewidth},frame=tb,basicstyle=\ttfamily,basicstyle=\footnotesize ]
(@\bf 1:@) a/source/org/jfree/chart/renderer/category/AbstractCategoryItemRenderer.java
(@\bf 2:@)         int index = this.plot.getIndexOf(this);
(@\bf 3:@)         CategoryDataset dataset = this.plot.getDataset(index);
(@{\bf 4:}@)         (@{\colorbox{codered}{{if (dataset != null) \{}}@)
(@\bf 5:@)             return result;
(@\bf 6:@)         }
\end{lstlisting}
    \caption{Buggy code fragment associated to patch in Fig.~\ref{fig:chart1}.}
    \label{fig:chart1-buggy}
\end{figure}

\vspace{-1mm}
\begin{figure}[!t]
    \centering
     \scriptsize
\begin{lstlisting}[language=Java,linewidth={\linewidth},frame=tb,basicstyle=\ttfamily, basicstyle=\footnotesize]
(@\bf 1:@) b/source/org/jfree/chart/renderer/category/AbstractCategoryItemRenderer.java
(@\bf 2:@)         int index = this.plot.getIndexOf(this);
(@\bf 3:@)         CategoryDataset dataset = this.plot.getDataset(index);
(@\bf 4:@)         (@{\colorbox{codegreen}{{if (dataset == null) \{}}@)
(@\bf 5:@)             return result;
(@\bf 6:@)         }
\end{lstlisting}
    \caption{Patched code fragment associated to patch in Fig.~\ref{fig:chart1}.}
    \label{fig:chart1-patched}
\end{figure}
%

\subsection{Embedding models}
\label{sec:embeddings}
\begin{figure}[!h]
	\includegraphics[width=0.89\columnwidth]{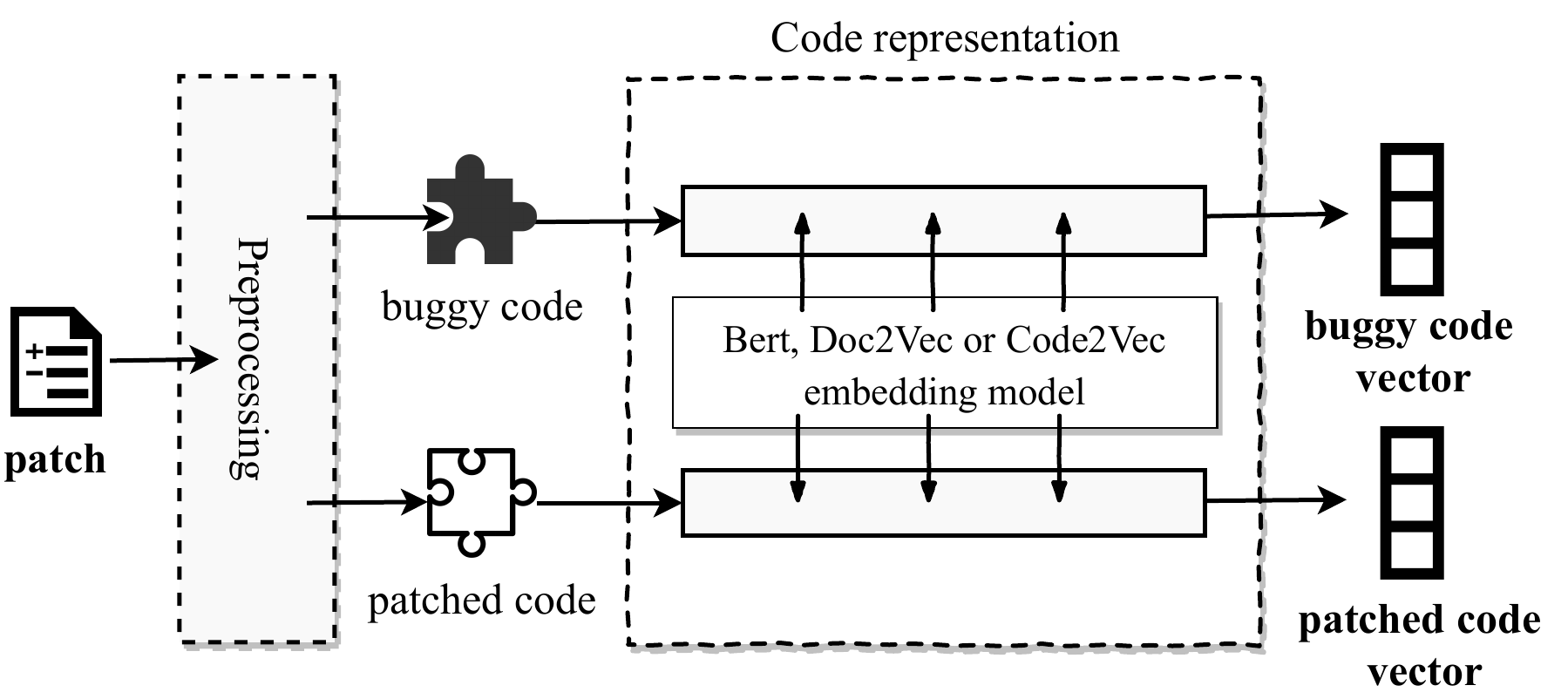}
	\caption{Producing code fragment embeddings with BERT, Doc2Vec and code2vec.}
	\label{fig:embedding}
\end{figure}

\vspace{-1mm}
When representation learning algorithms are applied to some training data, they produce {\em embedding models} that have learned to map a set of code tokens in the vocabulary of the training data to vectors of numerical values. These vectors are also referred to as {\em embeddings}.
Figure~\ref{fig:embedding} illustrates the process of embedding buggy code and patched code for the purpose of our experiments. 

The embedding models used in this work are obtained from different sources and training scenarios. 

\begin{figure}[!b]
	\includegraphics[width=0.89\columnwidth]{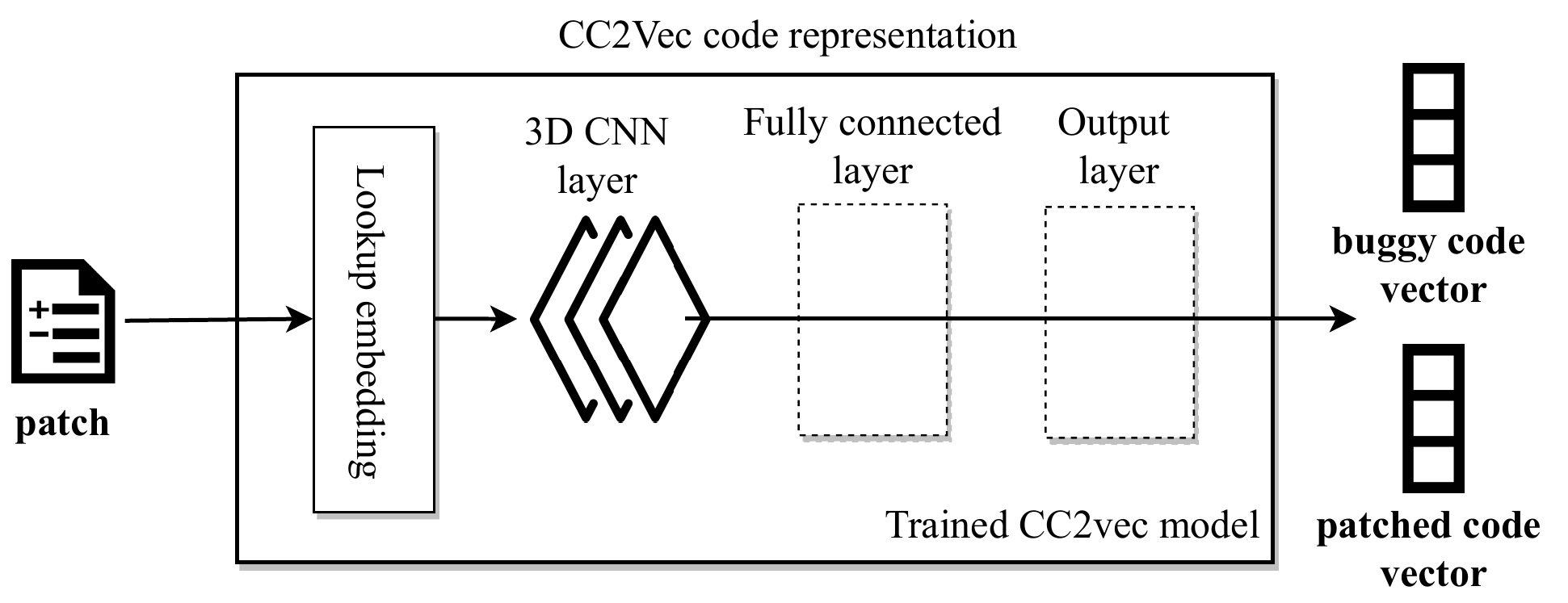}
	\caption{Extracting code fragment embeddings from CC2Vec pre-trained model.}
	\label{fig:cc2vec}
\end{figure}

\begin{itemize}[leftmargin=*]
	\item {\bf BERT.} In the first scenario, we consider an embedding model that initially targets natural language data, both in terms of the learning algorithm and in terms of training data. The network structure of BERT, however, is deep, meaning that it requires large datasets for training the embedding model. As it is now custom in the literature, we instead leverage a pre-trained 24-layer BERT model, which was trained on a Wikipedia corpus.
	\item {\bf Doc2Vec.} In the second scenario, we consider an embedding model that is trained on code data but using a representation learning technique that was developed for text data. To that end, we have trained the Doc2Vec model with code data of 36,364 patches from the 5 repair benchmarks (cf. Table~\ref{tab:datasets}).
	\item {\bf code2vec.} In the third scenario, we consider an embedding model that primarily targets code, both in terms of the learning algorithm and in terms of training data. We use in this case a pre-trained model of code2vec, which was trained by the authors using \textasciitilde 14 million code examples from Java projects. 
	\item {\bf CC2Vec.} Finally, in the fourth scenario, we consider an embedding model that was built in representation learning experiments for code changes. The pre-trained model that we leveraged from the work of Hoang~{\em et~al.}~\cite{hoang2020cc2vec} is embedding each patch into a single vector. We investigate the layers and identify the middle CNN-3D layer as the sweet spot to extract embeddings for buggy code and patched code fragments. Figure~\ref{fig:cc2vec} illustrates the process.
\end{itemize}


\section{Experiments }
\label{sec:eval}
We present the experiments that we designed to answer the research questions of our study. For each experiment, we state the objective, overview the execution details before presenting the results. 

\subsection{[Similarity Measurements for Buggy and Patched Code using Embeddings]}
\label{subsec:rq1}

{\bf Objective:} We investigate the capability of different learned embeddings to capture the similarity/dissimilarity between code fragments. The experiments are performed towards providing answers  for two sub-questions:
\begin{enumerate}
	\item[RQ-1.1] Is correct code actually similar to buggy code based on learned embeddings? 
	\item[RQ-1.2] To what extent is buggy code more similar to correctly-patched code than to incorrectly-patched code?
\end{enumerate}


{\bf Experimental Design:} We perform two distinct experiments with available datasets to answer RQ-1.1 and RQ-1.2. 

\noindent 
{\sc {\bf[Experiment \ding{182}]}} Using the four embedding models considered in our study (cf. Section~\ref{sec:embeddings}), we produce the embeddings for buggy and patched code fragments associated to 36k patches from five benchmark shown in Table~\ref{tab:data1}. In this case, the patched code fragment represents correct code since it comes from labeled benchmark data (generally representing developer fix patches). 
Given those embeddings (i.e., code representation vectors), we compute the cosine similarity between the vector representing the buggy code fragment and the vector representing the patched code fragment.

\begin{table}[!h]
	\centering
	\caption{Patch datasets used for computing similarity scores between buggy code fragments and correct code fragments.}
	\label{tab:data1}
	\resizebox{1\linewidth}{!}
	{
	\begin{threeparttable}
		\begin{tabular}{lccccc|c}
			 & {\bf \rotatebox[origin=c]{30}{Bears}} & {\bf \rotatebox[origin=c]{30}{\small Bugs.jar}} & {\bf \rotatebox[origin=c]{30}{\small Defects4J}} & {\bf \rotatebox[origin=c]{30}{\small ManySStuBs4J}} & {\bf \rotatebox[origin=c]{30}{\small QuixBugs}} & \rotatebox[origin=c]{30}{Total}\\
			\hline 
			\# Patches & 251 & 1,158 & 864 & 34,051 & 40 & 36,364$^\ast$\\
			\bottomrule
		\end{tabular}
	{$^\ast$Due to parsing failures, code2vec embeddings are available for 21,135 patches.}
	\end{threeparttable}
	}
\end{table}

\noindent
{\sc {\bf[Experiment \ding{183}]}} To compare the similarity scores of correct code fragment vs incorrect code fragment to the buggy code, we consider combining datasets with correct patches and datasets with incorrect patches. Note that, all patches in our experiments are plausible since we are focused on correctness: plausibility is straightforward to decide based on test suites. 
Correct patches are provided in benchmarks. However, incorrect patches associated to all benchmark bugs are not available. We rely on the dataset released by Liu~{\em et~al.}~\cite{liu2020efficiency}: 674 plausible but incorrect patches generated by 16 repair tools for 184 Defects4J bugs are considered from this dataset. Those 674 incorrect patches are selected within a larger set of incorrect patches by adding the constraint that the incorrect patch should be changing the same code location as the developer-provided patch in the benchmark: such incorrect patch cases may indeed be the most challenging to identify with heuristics.
 We propose to compare the similarity scores between the incorrect code and buggy code associated to  the dataset with the similarity scores between correct code and buggy associated to all benchmarks, all Defects4J benchmark data, or only the subset of Defects4J that corresponds to the 184 patches for which relevant incorrect patches are available.

\begin{figure}[!h]
\centering
	\includegraphics[width=1\linewidth]{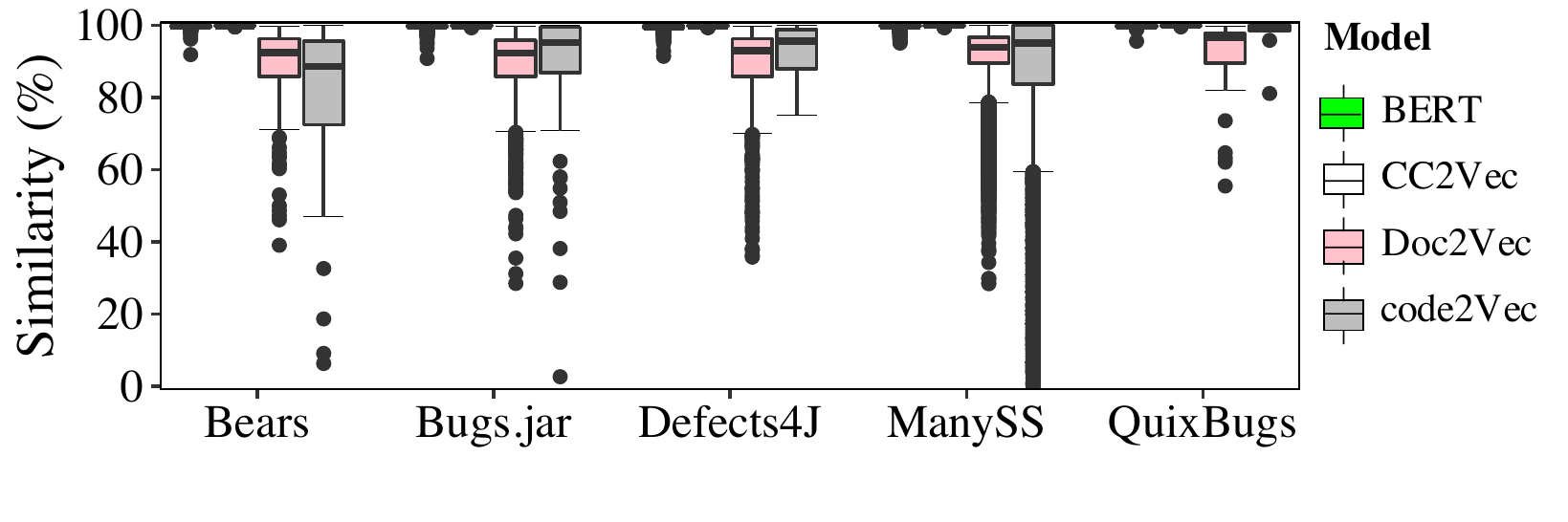}
	\vspace{-7.5mm}
	\caption{Distributions of similarity scores between correct and buggy code fragments. {\small ``ManySS'' denotes ``ManySStuBs4J''.}}
	\label{fig:all-sim-correct}
\end{figure}

{\bf Results:} 
Figure~\ref{fig:all-sim-correct} presents the boxplots of the similarity distributions with different embedding models and for samples in different datasets. Doc2Vec and code2vec models appear to yield similarity values that are lower than BERT and CC2Vec models.  

\begin{figure}[!t]
	\includegraphics[width=1\linewidth]{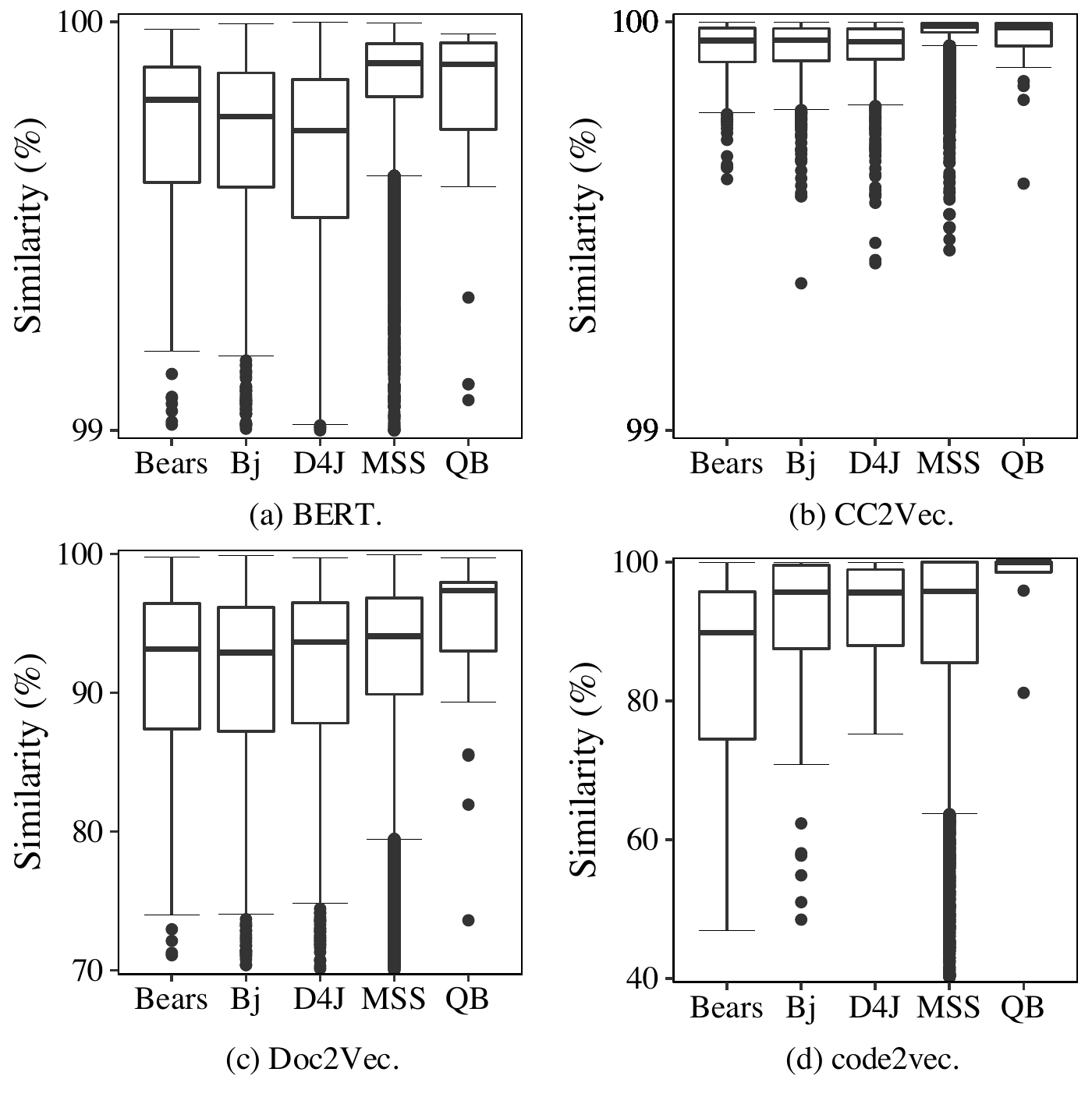}
	\vspace{-5mm}
	\caption{Zoomed views of the distributions of similarity scores between correct and buggy code fragments.}
	\vspace{1mm}
	\label{fig:results-1}
\end{figure}

\vspace{-1mm}
Figure~\ref{fig:results-1} zooms in the boxplot region for each embedding model experiment to overview the differences across different benchmark data.
We obverse that, when embedding the patches with BERT, the similarity distribution for the patches in Defects4J dataset is similar to Bugs.jar and Bears dataset, but is different from the dataset ManySStBs4J and QuixBugs. The Mann-Whitney-Wilcoxon (MWW) tests~\cite{wilcoxon1945individual,mann1947test} confirm that the similarity of median scores for Defects4J, Bugs.jar and Bears is indeed statistically significant. MWW tests further confirms the statistical significance of the difference between Defects4J and ManySStBs4J/QuixBugs scores.


Defects4J, Bugs.jar and Bears include diverse human-written patches for a large spectrum of bugs from real-world open-source Java projects. In contrast, ManySStuBs4J only contains patches for single statement bugs. Quixbugs dataset is limited by its size and the fact that the patches are built by simply mutating the code of small Java implementation of 40 algorithms (e.g., sort, sieve, etc.).

While CC2Vec and Doc2Vec exhibit roughly similar patterns with BERT (although at different scales), the experimental results with code2vec present different patterns across datasets. Note that, due to parsing failures of code2vec, we eventually considered only 118 Bears patches, 123 Bugs.jar patches, 46 Defects4J patches, 20,840 ManySStuBs4J patches and 8 QuixBugs. The change of dataset size could explain the difference with the other embedding models.

\find{{\bf \ding{45} RQ1.1 }$\blacktriangleright$  Learned representations of buggy and correct code fragments exhibit high cosine similarity scores. Median scores are similar for patches that are collected with similar heuristics (e.g., in the wild patches vs single-line patches vs debugging example patches). The pre-trained BERT natural language model captures more similarity variations than the CC2Vec model, which is specialized for code changes.$\blacktriangleleft$ }

\begin{figure}[!ht]
	\includegraphics[width=1\linewidth]{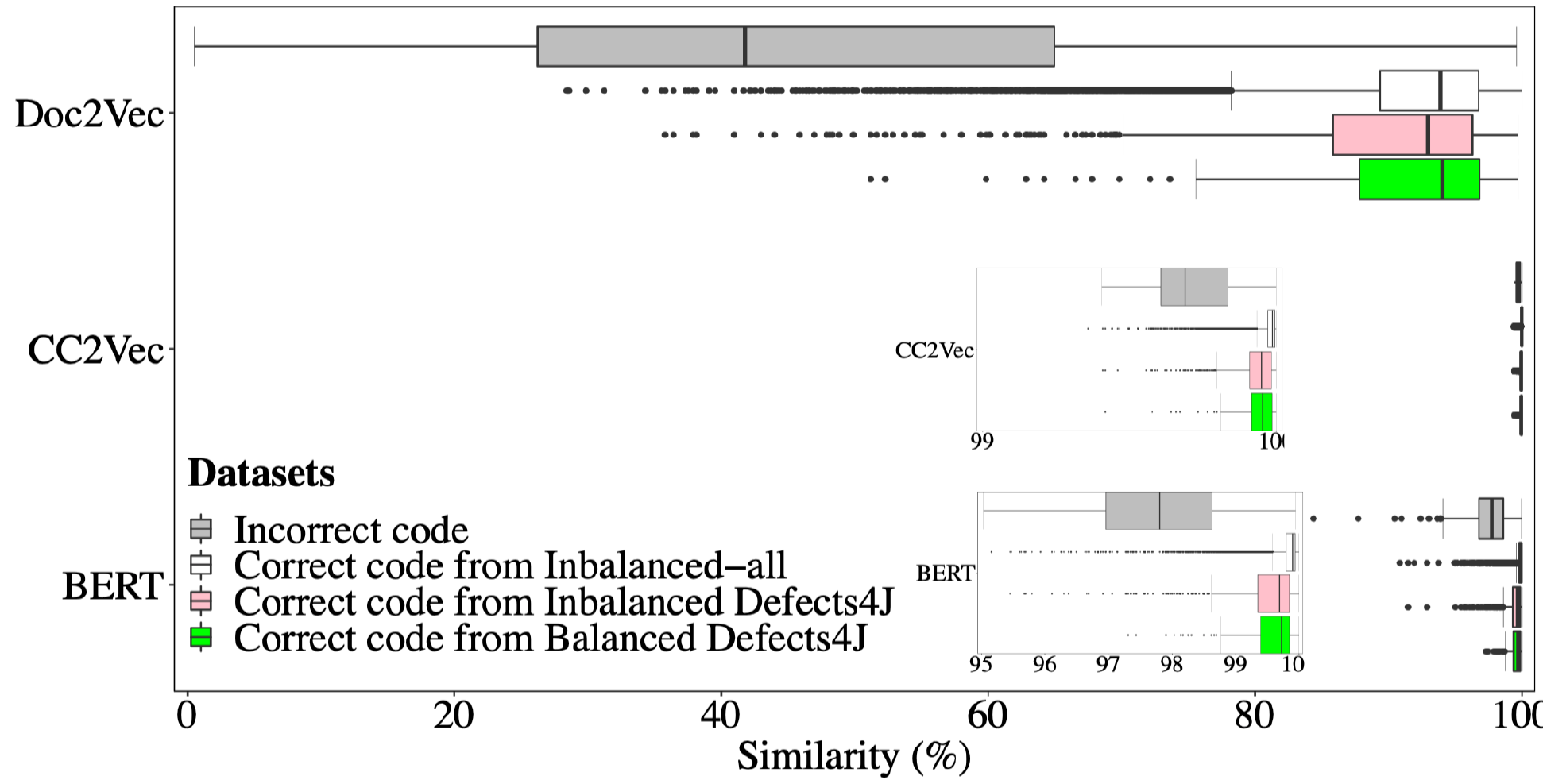}
	\caption{Comparison of similarity score distributions for code fragments in incorrect and correct patches.}
	\label{fig:c-vs-i-3}
\end{figure}

\vspace{-1.5mm}
In the second experiment, we further assess whether incorrectly-patched code exhibits different similarity score distributions than correctly-patched code. Figure~\ref{fig:c-vs-i-3} shows the distributions of cosine similarity scores for correct patches (i.e., similarity between buggy code and correct code fragments) and incorrect patches (i.e., similarity between buggy code and incorrect code fragments). The comparison is done with different scenarios specified in Table~\ref{tab:data2}.

\begin{table}[!h]
	\centering
	\caption{Scenarios for similarity distributions comparison.}
	\label{tab:data2}
	\resizebox{1\linewidth}{!}
	{
	\begin{threeparttable}
		\begin{tabular}{l|p{30mm}|L{40mm}}
			\toprule
			{\bf Scenario} & {\bf Incorrect patches} & {\bf Correct patches}\\
			\hline
			Imbalanced-al$^\ast$ & 674 incorrect patches 
		  &  all correct patches from 5 benchmarks in Table~\ref{tab:data1}.  \\\cline{1-1}\cline{3-3}
			Imbalanced-Defects4J & by 16 APR tools~\cite{liu2020efficiency}  & all correct patches from Defects4J.  \\\cline{1-1}\cline{3-3}
			Balanced-Defects4J & for 184 Defects4J bugs& all correct patches for the 184 Defects4J bugs.  \\
			\bottomrule
		\end{tabular}
		{$^\ast$Except for Defects4J, there are no publicly-released incorrect patches for APR datasets.}
	\end{threeparttable}
	}
\end{table}

The comparisons do not include the case of embeddings for code2vec. 
Indeed, unlike the previous experiment where code2vec was able to parse enough code fragments, for the considered 184 correct patches of Defects4J, code2vec failed to parse most of the relevant code fragments. Hence, we focus the comparison on the other three embedding models (BERT, Doc2Vec and CC2Vec). Overall, we observe that the distribution of cosine similarity scores is substantially different for correct and incorrect code.

We observe that the similarity distributions of buggy code and patched code from incorrect patches are significantly different from the similarities for correct patches. The difference of median values is confirmed to be statistically significant by an MWW test.
Note that the difference remains high for BERT, Doc2Vec and CC2Vec whether the correct code is the counterpart of the incorrect ones (i.e., the scenario of Balanced-Defects4J) or whether the correct code is from a larger dataset (i.e., Imbalanced-all and Imbalanced-Defects4J scenarios).

\find{{\bf \ding{45} RQ1.2 }$\blacktriangleright$ Learned representations of code fragments with BERT, CC2Vec and Doc2Vec yield similarity scores that, given a buggy code, substantially differ between correct code and incorrect code. This result suggests that similarity score can be leveraged to discriminate correct patches from incorrect patches.$\blacktriangleleft$}
\vspace{-3mm}
\subsection{[Filtering of Incorrect Patches based on Similarity Thresholds]}
\label{subsec:rq2}

{\bf Objective:} Following up on the findings related to the first research question, we investigate the selection of cut-off similarity scores to decide on which APR-generated patches are likely incorrect. 
Results from this investigation will provide insights to guide the exploitation of code embeddings in program repair pipelines.
 
{\bf Experimental design:}
To select threshold values, we consider the distributions of similarity scores from the above experiments (cf. Section~\ref{subsec:rq1}). Table~\ref{tab:sim2} summarizes relevant statistics on the distributions on the similarity scores distribution for correct patches. Given the differences that were exhibited with incorrect patches in previous experiments, we use, for example, the 1$^{st}$ quartile value as an inferred threshold value.  

\begin{table}[!h]
	\centering
	\caption{Statistics on the distributions of similarity scores for correct patches of Bears+Bugs.jar+Defects4J.}
	\label{tab:sim2}
	\resizebox{1\linewidth}{!}
	{
	\begin{threeparttable}
		\begin{tabular}{l|cccccc}
			\toprule
			{\bf Subjects} & {\bf Min.} & {\bf 1st Qu.} & {\bf Median} & {\bf 3rd Qu.} & {\bf Max.} & {\bf Mean}\\
			\hline
			BERT   & 90.84 & 99.47 & 99.73 & 99.86 & 100 & 99.54 \\
			CC2Vec & 99.36 & 99.91 & 99.95 & 99.98 & 100 & 99.93 \\
			Doc2Vec& 28.49 & 85.80 & 92.60 & 96.10 &99.89& 89.19 \\
			code2vec& 2.64 & 81.19 & 93.63 & 98.87 & 100 & 87.11 \\
			\bottomrule
		\end{tabular}
	\end{threeparttable}
	}
\end{table}

Given our previous findings that different datasets exhibit different similarity score distributions, we also consider inferring a specific threshold for the QuixBugs dataset (cf. statistics in  Table~\ref{tab:sim3}). We do not compute any threshold based on ManySStuBs4J since it has not yet been applied to program repair tools.

\begin{table}[!h]
	\centering
	\caption{Statistics on the distributions of similarity scores for correct patches of QuixBugs.}
	\label{tab:sim3}
	\resizebox{1\linewidth}{!}
	{
	\begin{threeparttable}
		\begin{tabular}{l|cccccc}
			\toprule
			{\bf Subjects} & {\bf Min.} & {\bf 1st Qu.} & {\bf Median} & {\bf 3rd Qu.} & {\bf Max.} & {\bf Mean}\\
			\hline
			BERT   & 95.63 & 99.69 & 99.89 & 99.95 & 99.97 & 99.66 \\
			CC2Vec & 99.60 & 99.94 & 99.99 & 100   & 100   & 99.95 \\
			Doc2Vec& 55.51 & 89.56 & 96.65 & 97.90 & 99.72 & 91.29 \\
			code2vec& 81.16& 98.	53 & 100   & 100   & 100   & 97.06 \\
			\bottomrule
		\end{tabular}
	\end{threeparttable}
	}
\end{table}

\begin{table*}[!t]
	\centering
	\caption{Filtering incorrect patches by generalizing thresholds inferred from Section~\ref{subsec:rq1}.Results.}
	\label{tab:filtering}
	\resizebox{1\textwidth}{!}
	{
	\begin{threeparttable}
		\begin{tabular}{l|cc|c|cccc|cccc|cccc}
			\toprule
			\multirow{2}{*}{\bf Dataset} & \multirow{2}{*}{\bf \# CP} & \multirow{2}{*}{\bf \# IP} &\multirow{2}{*}{\bf Threshold} & \multicolumn{4}{c|}{\bf BERT} & \multicolumn{4}{c|}{\bf CC2Vec} & \multicolumn{4}{c}{\bf Doc2Vec}\\\cline{5-16}
			 & & & &{\bf \# +CP} & {\bf \# -IP} & {\bf +Recall}& {\bf -Recall}& {\bf \# +CP} & {\bf \# -IP}& {\bf +Recall}& {\bf -Recall}& {\bf \# +CP} & {\bf \# -IP}& {\bf +Recall}& {\bf -Recall}\\
			\hline
			\multirow{2}{*}{\makecell[l]{Bears, Bugs.jar\\ and Defects4J}}&\multirow{2}{*}{893} &\multirow{2}{*}{61,932}& 1st Qu.&57& 48,846 & 6.4\% & \cellcolor{black!25}78.9\%&797&19,499 & \cellcolor{black!25}89.2\% & 31.5\%&794&25,192 & 88.9\% & 40.7\%\\\cline{4-16}
			& & & Mean & 49&51,783 & 5.5\% & \cellcolor{black!25}83.6\%&789&23,738 & \cellcolor{black!25}88.4\%& 38.3\%&771&33,218&86.3\%& 53.6\% \\	\hline\hline
			\multirow{2}{*}{QuixBugs}&\multirow{2}{*}{7} &\multirow{2}{*}{1,461} & 1st Qu. & 4 & 1,387 & 57.1\% & \cellcolor{black!25}94.9\% & 4 & 1,198& 57.1\% & 82.0\%& 7 & 1,226 & \cellcolor{black!25}100\% & 83.9\%\\\cline{4-16}
			& & & Mean & 4 & 1,378 &57.1\%& \cellcolor{black!25}94.3\% &  4 & 1,255&57.1\%&85.9\% &  7 & 1270& \cellcolor{black!25}100\%&86.9\% \\		
			\bottomrule
		\end{tabular}
		{$^\ast$``{\bf \# CP}'' and ``{\bf \# IP}'' stand for the number of correct and incorrect patches, respectively. ``{\bf \# +CP}'' means the number of correct patches that can be ranked upon the threshold, while ``{\bf \# -IP}'' means the number of incorrect patches that can be filtered out by the threshold. ``{\bf +Recall}'' and ``{\bf -Recall}'' represent the recall of identifying correct patches and filtering out incorrect patches, respectively.}
	\end{threeparttable}
	}
\end{table*}

Our test data is constituted of 64,293 patches generated by 11 APR tools in the empirical study of Durieux~{\em et~al.}~\cite{durieux2019empirical}. First, we use the four embedding models to generate embeddings of buggy and patched code fragments and compute cosine similarity scores. Second, for each bug, we rank all generated patches based on the similarity score between the patched code and the buggy, where we consider that the higher the score, the more likely the correctness. 
Finally, to filter incorrect candidates, we consider two experiments: 
\begin{enumerate}[leftmargin=*]\vspace{-1mm}
\item Patches that lead to similarity scores that are lower to the inferred threshold (i.e., 1$^{st}$ Quartile in previous experimental data) will be considered as incorrect. Patches where patched code exhibit higher similarity scores than the threshold are considered likely correct. 
\item Another approach is to consider only the top-1 patches with the highest similarity scores as correct patches. Other patches are considered incorrect.
\end{enumerate}

\vspace{-1mm}
In all cases, we systematically validate the correctness of all 64,293 patches to have the correctness labels, for which the dataset authors did not provide (all plausible patches having been considered as valid). First, if the file(s) modified by a patch are not the same buggy files in the benchmark, we systematically consider it as incorrect: with this simple scheme, 33\,489 patches are found incorrect. Second, with the same file, if the patch is not making changes at the same code locations, we consider it to be incorrect: 26\,386 patches are further tagged as incorrect with this decision (cf. Threats to validity in Section~\ref{sec:dis}). Finally, for the remaining 4\,418 plausible patches in the dataset, we manually validate  correctness by following the strict criteria enumerated by Liu~{\em et~al.}~\cite{liu2020efficiency} to enable reproducibility. Overall, we could label 900 correct patches. The remainders are considered as incorrect.

%


{\bf Results:} By considering the patch with the highest (top-1) similarity score between the patched code and buggy code as correct, we were able to identify a correct patch for 10\% (with BERT), 9\% (with CC2Vec) and 10\% (with Doc2Vec) of the bug cases. Overall we also misclassified 96\% correct patches as incorrect. However, only 1.5\% of incorrect patches were misclassified as correct patches.

Given that a given bug can be fixed with several correct patches, the top-1 criterion may not be adequate. Furthermore, this criterion makes the assumption that a correct patch indeed  exists among the patch candidates. By using filtering thresholds inferred from previous experiments (which do not include the test dataset in this experiment), we can attempt to filter all incorrect patches generated by APR tools. Filtering results presented in Table~\ref{tab:filtering} show the recall scores that can be reached. We provide experimental results when we use 1$^{st}$ Quartile and Mean values of similarity scores in the "training" set as threshold values. The threshold are also applied by taking into account the datasets: thresholds learned on QuixBugs benchmark are applied to generated patches for QuixBugs bugs. 

\find{{\bf \ding{45} RQ2 }$\blacktriangleright$Building on cosine similarity scores, code fragment embeddings can help to filter out between 31.5\% with CC2Vec and 94.9\% with BERT of incorrect patches. While BERT achieves the highest recall of filtering incorrect patches, it produces embeddings that lead to a lower recall (at 5.5\%) at identifying correct patches.$\blacktriangleleft$}
\subsection{[Classification of Correct Patches with supervised learning]}

{\bf Objective:} Cosine similarity between embeddings (which was used in the previous experiments) considers every deep learned feature as having the same weight as the others in the embedding vector. We investigate the feasibility to infer, using machine learning, the weights that different features may present with respect to patch correctness. We compare the prediction evaluation results with the achievements of related approaches in the literature.

{\bf Experimental design:} To perform our machine learning experiments, we first require a ground-truth dataset. To that end, we rely on labeled datasets in the literature. Since incorrect patches generated by actual APR tools are only available for the Defects4J bugs, we focus on labeled patches provided by two independent teams (Liu~{\em et~al.}~\cite{liu2020efficiency} and Xiong~{\em et~al.}~\cite{xiong2018identifying}). Very few patches generated by the different tools are actually labeled as correct, leading to an imbalanced dataset. To reduce the imbalance issue, we supplement the dataset with developer (correct) patches as supplied in the Defects4J benchmark. Eventually, our dataset shown in Table~\ref{tab:data3} included 1000 patches after removing duplicates to avoid data bias.

\begin{table}[!ht]
	\centering
	\caption{Dataset for evaluating ML-based predictors of patch correctness.}
	\label{tab:data3}
	\resizebox{1\linewidth}{!}
	{
	\begin{threeparttable}
		\begin{tabular}{l|c|c|r}
			\toprule
			& {\bf Correct patches}& {\bf Incorrect patches} & {\bf Total}\\
			\hline
			 Liu~{\em et~al.}~\cite{liu2020efficiency} &137& 502&639\\
			 Xiong~{\em et~al.}~\cite{xiong2018identifying} &30 &109 &139\\
			 Defects4J (developers)~\cite{just2014defects4j}&356 &0 & 356\\ \hline
						 Whole dataset  & 523&611 &1134\\
			\bottomrule
			\bf Final Dataset (deduplicated) & \bf 468 &\bf 532 &\bf 1000\\
			\bottomrule
		\end{tabular}
	\end{threeparttable}
	}
\end{table} 

Our ground truth dataset patches  are then fed to our embedding models to produce embedding vectors. As for previous experiments, the parsability of Defects4J patch code fragments prevented the application of code2vec: we use pre-trained models of BERT (trained with natural language text) and CC2Vec (trained with code changes) as well as a retrained model of Doc2Vec (trained with patches). 

\begin{figure}[!t]
	\includegraphics[width=1\columnwidth]{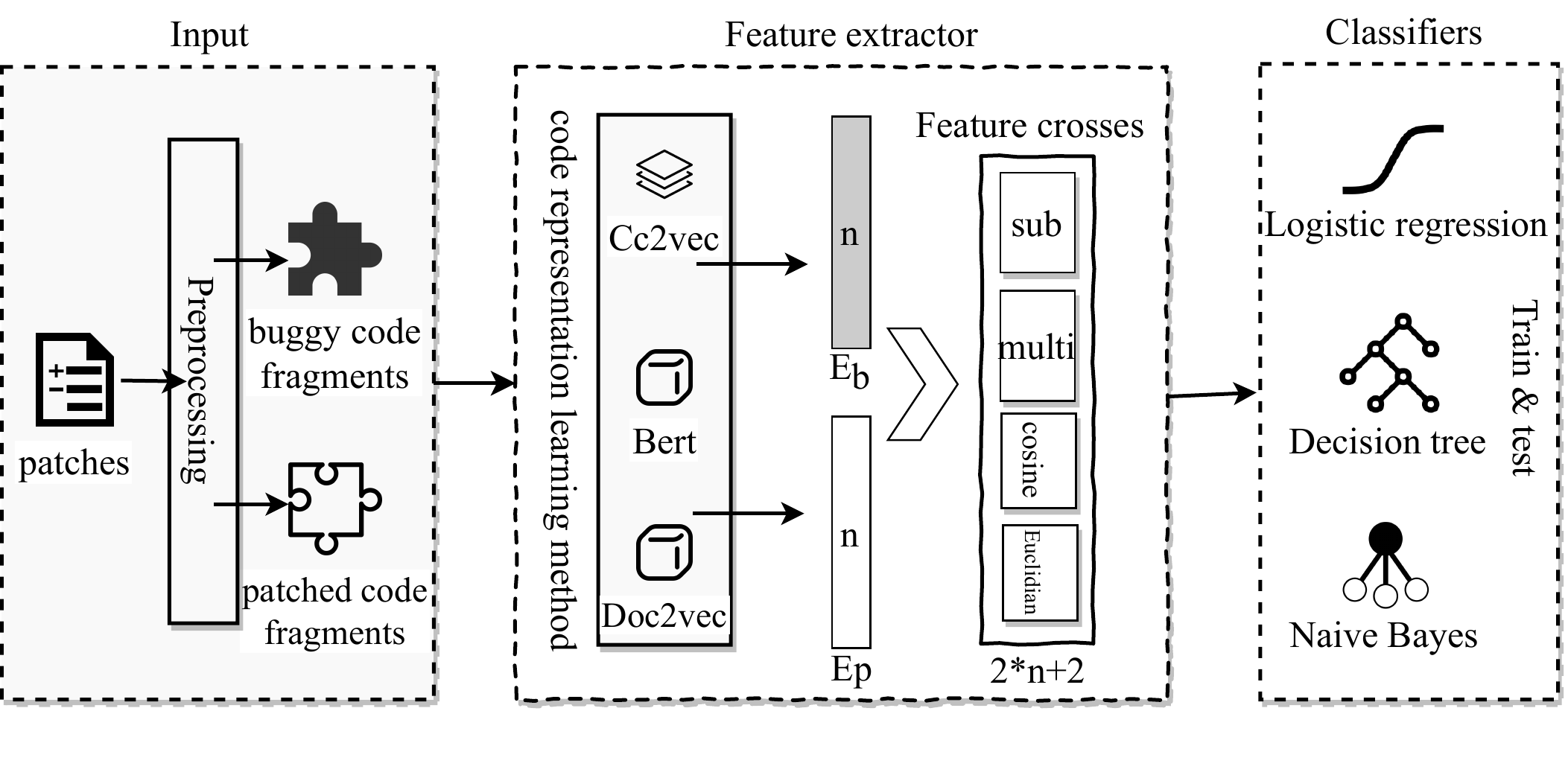}
	\vspace{-7mm}
	\caption{Feature engineering for correctness classification.}
	\label{fig:pipeline}
\end{figure}

Since the representation learning models are applied to code fragments inferred from patches (and not to the patch themselves), we collect the embeddings of both buggy code fragment and patched code fragment for each patch. Then we must merge these vectors back into a single input vector for the classification algorithm. We follow an approach that was demonstrated by Hoang et al.~\cite{hoang2020cc2vec} in a recent work on bug fix patch prediction: the classification model performs best when features of patched code fragment and buggy code fragment are crossed together. We thus propose a classification pipeline (cf. Figure~\ref{fig:pipeline}) where the feature extraction for a given patch is done by applying subtraction, multiplication, cosine similarity and euclidean similarity to capture crossed features between the buggy code vector and the patched code vector. The resulting patch embedding has 2*n+2 dimensions where n is the dimension of input code fragment embeddings. 

{\bf Results:} We compare the performance of different predictors (varying the embeding models) using different learners (i.e., classification algorithms). Results presented in Table ~\ref{tab:ML-all} are averaged from a 5-fold cross validation setup. All classical metrics used for assessing predictors are reproted: Accuracy, Precision, Recall, F1-Measure, Area Under Curve (AUC). Logistic Regression (LR) applied to BERT embeddings yield the best performance measurements: 0.720 for F1 and 0.808 for AUC.

\begin{table}[!h]
	\centering
	\caption{Evaluation on three ML classifiers.}
	\label{tab:ML-all}
	\resizebox{1\linewidth}{!}
	{
	\begin{threeparttable}
		\begin{tabular}{l|c|ccccc}
			\toprule
			{\bf Classifier} & {\bf Embedding} &{\bf Acc.} & {\bf Prec.} & {\bf Recall.} & {\bf F1} & {\bf AUC} \\
			\hline
			\multirow{3}{*}{DecisionTree} & BERT  & 63.6 & 62.0 & 57.3 & 59.6 & 0.632 \\
			& CC2Vec  & 69.0 & 66.9 & 68.0 & 67.2 & 0.690 \\
			& Doc2Vec & 60.2 & 57.4 & 57.7 & 57.5 & 0.600 \\
			\hline
			\multirow{3}{*}{Logistic regression} & BERT& 74.4 & 73.8 & 70.3 & \cellcolor{black!25}72.0 & \cellcolor{black!25}0.808 \\
			& CC2Vec  & 73.9 & 72.5 & 72.0 & 72.0 & 0.788 \\
			& Doc2Vec & 66.3 & 65.3 & 59.9 & 62.3 & 0.707 \\
			\hline
			\multirow{3}{*}{Naive bayes} & BERT& 60.3 & 55.6 & 77.0 & 64.5 & 0.642 \\
			& CC2Vec  & 58.0 & 65.4 & 22.7 & 28.5 & 0.722 \\
			& Doc2Vec & 66.3 & 69.4 & 49.8 & 57.9 & 0.714 \\
			\bottomrule
		\end{tabular}
	\end{threeparttable}
	}
\end{table}

\vspace{-1mm}
\find{{\bf \ding{45} RQ3.1 }$\blacktriangleright$ An ML classifier trained using Logistic Regression with BERT embeddings yield very promising performance on patch correctness prediction (F-Measure at 72.0\% and AUC at 80.8\%). $\blacktriangleleft$}
\vspace{-1mm}

{\sc [Comparison against the state of the art].} There are two related works for patch prediction which were both evaluated on 139 patches released by Xiong~{\em et~al.}~\cite{xiong2018identifying}. PATCH-SIM~\cite{xiong2018identifying} compares execution traces of patched programs to identify correctness. ODS~\cite{ye2019automated} leverages manually-crafted features to build machine learning classifiers. 

\begin{figure}[!t]
	\includegraphics[width=0.7\columnwidth]{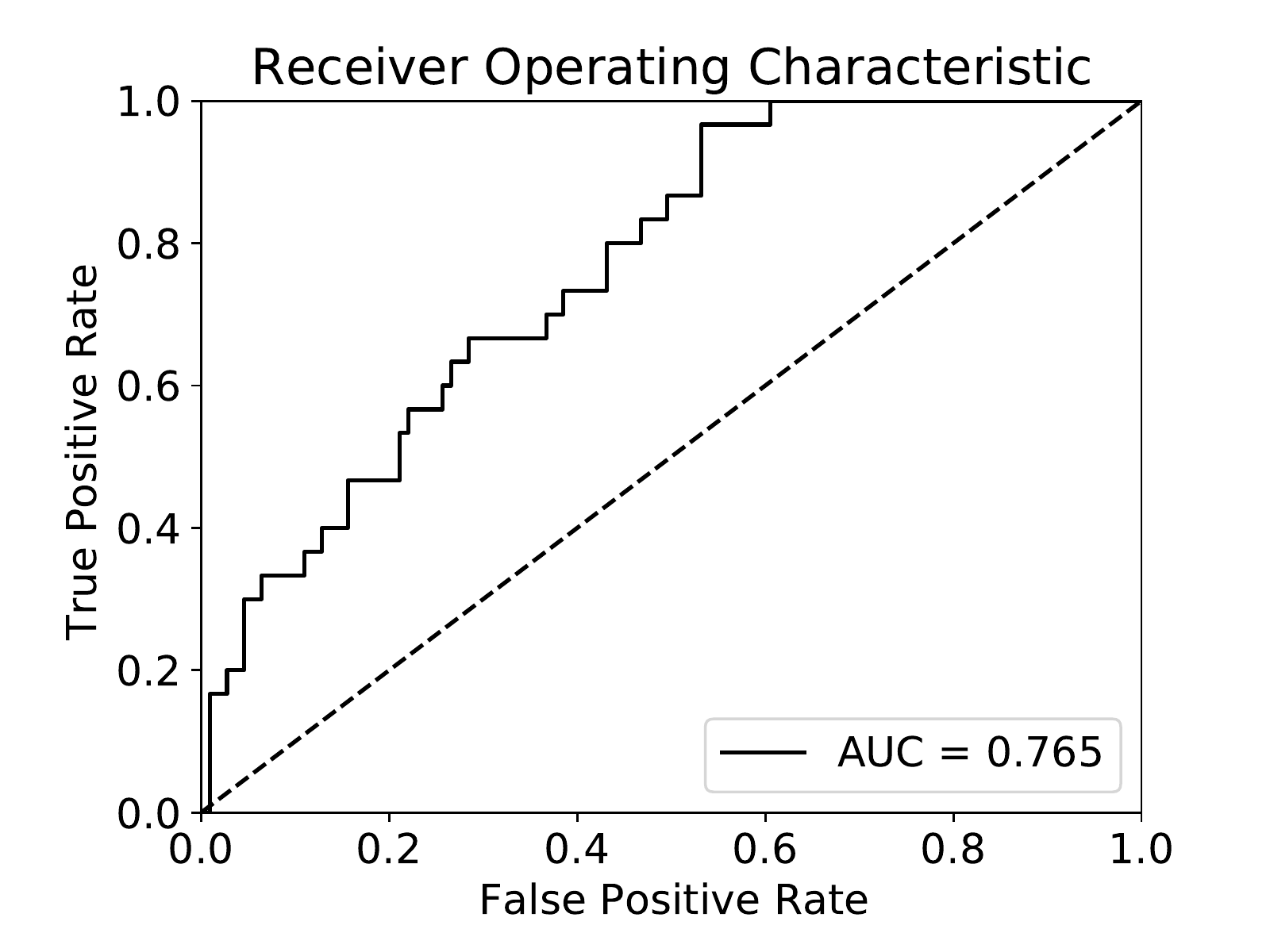}
	\caption{Performance of ML patch correctness predictor using BERT/Logistic Regression: Test set from~\cite{xiong2018identifying}.}
	\label{fig:ROC}
\end{figure}

We consider the 139 patches as test set and the remainder in our dataset ($870=1000-130$\footnote{9 patches in the ground truth dataset by Xiong et al.~\cite{xiong2018identifying} were duplicates (e.g., Patch151 $\equiv$ Patch23).}) for training. Note that the 139 patches are associated to bug cases where repair tools can generate patches. These patches may thus be substantially different from the rest in our dataset. Indeed our best learner (Logistic Regression with BERT embeddings) yields an AUC of 0.765.
The Receiver Operating Characteristic (ROC) curve is presented in Figure~\ref{fig:ROC}. 

In the validation of PATCH-SIM~\cite{xiong2018identifying}, the authors aimed for avoiding to filter out any correct patches. Eventually, when guaranteeing that no correct patch is excluded, they could still exclude  62 (56.3\%) incorrect patches. If we constrain the threshold of our predictor to avoid misclassifying any correct patch (threshold value = 0.219), our predictor is able to exclude up to 43 (39.4\%)  incorrect patches, which represents a reasonably promising achievement since no dynamic information is used (in contrast to PATCH-SIM). Table~\ref{tab:dyn-stat} overviews the prediction results comparison.

\begin{table}[!h]
	\centering
	\caption{Comparison of incorrect patch identification between PATCH-SIM (uses dynamic information) and BERT+LR (uses embeddings statically inferred from patches).}
	\label{tab:dyn-stat}
	\resizebox{1\linewidth}{!}
	{
	\begin{threeparttable}
		\begin{tabular}{l|cc|cc|cc}
			\toprule
			& \multicolumn{2}{c}{Ground Truth} & \multicolumn{2}{c}{PATCH-SIM} & \multicolumn{2}{c}{BERT + LR} \\
			\toprule
			{\bf Project} & {\bf \rotatebox[origin=c]{0}{Incorrect}} & {\bf \rotatebox[origin=c]{0}{Correct}} & {\bf \rotatebox[origin=c]{0}{Incorrect}} & {\bf  \rotatebox[origin=c]{0}{Correct}} & {\bf \rotatebox[origin=c]{0}{Incorrect}} & {\bf  \rotatebox[origin=c]{0}{Correct}}  \\
			 & {\bf \rotatebox[origin=c]{0}{}} & {\bf \rotatebox[origin=c]{60}{}} & {\bf \rotatebox[origin=c]{0}{excluded (\%)}} & {\bf  \rotatebox[origin=c]{0}{excluded}} & {\bf \rotatebox[origin=c]{0}{excluded (\%)}} & {\bf  \rotatebox[origin=c]{0}{excluded}}  \\
			\hline
			Chart   & 23 & 3 & 14(60.9\%) & 0&16(69.6\%) & 0 \\
			Lang   & 10 & 5 & 6(54.5\%) & 0&1(10\%) & 0 \\
			Math   & 63 & 20 & 33(52.4\%) &0& 23(36.5\%) & 0 \\
			Time   & 13 & 2 & 9(69.2\%) & 0&3(23.1\%) & 0 \\
			\midrule
			Total   & 109 & 30 & \cellcolor{black!25}62(56.3\%) &0& \cellcolor{black!25}43(39.4\%) & 0 \\
			\bottomrule
		\end{tabular}
	\end{threeparttable}
	}
\end{table}

We also compare the predictive power of our models against that of ODS~\cite{ye2019automated}, which builds on manually engineered features. We directly compare against the results reported by the authors on the 139 test patches. While the pre-trained BERT model associated with Logistic Regression (LR) achieves better AUC than ODS LR-based model (0.765 vs 0.705), ODS Random Forest-based model achieves a  higher AUC at 0.841. Note however that ODS has been trained on over 13 thousand patches (including patches for bugs associated to the test set patch), our training dataset includes only 870 patches (i.e., $\sim$1/20$^{th}$ of their dataset).

Tables~\ref{tab:bert-matrix} and~\ref{tab:ods-matrix}  provide confusion matrices for different cut-off thresholds of the classifiers for ODS and    our BERT embeddings-based classifiers: TP (true positives) represent correct patches that were classified as such; TN (true negatives) represent incorrect patches that were classified as such; FP (false positives) represent incorrect patches that were classified as correct; and FN (false negatives) represent correct patches that were classified as incorrect. Overall, the BERT-based predictor is very sensitive to the cut-off thresholds while ODS is less sensitive. We also note that BERT embeddings applied to Random Forrest does not yield good performance: decision trees are indeed known to be good for categorical data and request large datasets for training. In our case, the data set is small, while ODS has a training dataset that is about 20 times larger. The hand-crafted features of ODS may also help split the patches into categories while our deep learned features are based on a large vocabulary of natural language text. 

\begin{table}[!t]
	\centering
	\caption{Confusion matrix of ML predictions based on BERT embedddings with different thresholds.}
	\label{tab:bert-matrix}
	\resizebox{0.9\linewidth}{!}
	{
	\begin{threeparttable}
		\begin{tabular}{lc|cccccccccc}
			\toprule
			\multirow{2}{*}{\bf Learners} & \multirow{2}{*}{\bf AUC} &  \multicolumn{10}{c}{\bf Thresholds} \\\cline{3-12}
			 & & & 0.1 & 0.2 & 0.3 & 0.4 & 0.5 & 0.6 & 0.7 & 0.8 & 0.9\\
			\hline
			\multirow{4}{*}{LR}&\multirow{4}{*}{0.765} & \#TP & 30 & 30 & 24 & 19 & 16 & 12 & 10 & 6 & 4 \\\cline{3-12}
			 & & \#TN & 13 & 37 & 61 & 79 & 85 & 95 & 100 & 106 & 108 \\\cline{3-12}
			 & & \#FP & 96 & 72 & 48 & 30 & 24 & 14 & 9 & 3 & 1 \\\cline{3-12}
			 & & \#FN & 0 & 0 & 6 & 11 & 14 & 18 & 20 & 24 & 26 \\\hline
			 \multirow{4}{*}{RF}&\multirow{4}{*}{0.751} & \#TP & 30 & 30 & 29 & 26 & 20 & 12 & 4 & 2 & 0 \\\cline{3-12}
			 & & \#TN & 1 & 1 & 6 & 32 & 79 & 102 & 107 & 108 & 109 \\\cline{3-12}
			 & & \#FP & 108 & 108 & 103 & 77 & 30 & 7 & 2 & 1 & 0 \\\cline{3-12}
			 & & \#FN & 0 & 0 & 1 & 4 & 10 & 18 & 26 & 28 & 30 \\

			\bottomrule
		\end{tabular}
	\end{threeparttable}
	}
\end{table}

\begin{table}[!t]
	\centering
	\caption{Confusion matrix of ODS predictions with different thresholds.}
	\label{tab:ods-matrix}
	\resizebox{0.9\linewidth}{!}
	{
	\begin{threeparttable}
		\begin{tabular}{lc|cccccccccc}
			\toprule
			\multirow{2}{*}{\bf Learners} & \multirow{2}{*}{\bf AUC} &  \multicolumn{10}{c}{\bf Thresholds} \\\cline{3-12}
			 & & & 0.1 & 0.2 & 0.3 & 0.4 & 0.5 & 0.6 & 0.7 & 0.8 & 0.9\\
			\hline
			\multirow{4}{*}{LR}&\multirow{4}{*}{0.705} & \#TP &27 &27 &27 &27 &27 &27 &27 &27 &27 \\\cline{3-12}
			 & & \#TN &50 & 50& 50& 50& 50& 51& 51&52 &52 \\\cline{3-12}
			 & & \#FP & 60& 60& 60& 60& 60&59 &59 &58 &58 \\\cline{3-12}
			 & & \#FN & 2& 2& 2& 2& 2& 2& 2& 2&2 \\\hline
			 \multirow{4}{*}{RF}&\multirow{4}{*}{0.841} & \#TP & 29&29 &29 &29 &29 & 29& 25&23&14 \\\cline{3-12}
			 & & \#TN & 20&33 &36 &43 &51 &60 &68 &81 &101 \\\cline{3-12}
			 & & \#FP & 90& 77& 74& 67& 59& 50& 42& 29& 13\\\cline{3-12}
			 & & \#FN &0 &0 &0 &0 &0 &0 &4 &6 &15 \\
			\bottomrule
		\end{tabular}
	\end{threeparttable}
	}
\end{table}

We observe nevertheless that LR classifiers fed with BERT embeddings are able to recall high numbers of incorrect patches (\#TN is high and \#FP is low on threshold > 0.5). In contrast ODS consistently recalls correct patches (however with high false positives).  
These experimental results suggest that both approaches can be used in a complementary way. In future work, we will propose an approach that carefully merges deep learned features to hand-crafted features towards yielded a better predictors of patch correctness.

\find{{\bf \ding{45} RQ3.2 }$\blacktriangleright$ML predictors trained on learned representations appear to perform slightly less well than state of the art PATCH-SIM approach which relies on dynamic information. On the other hand, deep code representations appear to be complementary to hand-crafted features engineered for ODS. Overall, we recall that our experimental evaluations are performed in a zero-shot scenario, i.e., without fine-tuning the parameters of any of the pre-trained models. Furthermore, the training dataset of the classifiers is an order of magnitude smaller\footnote{We were not able to collect or reconstitute the training dataset used in ODS to train our model.} than the one used by most closely-related work (i.e., ODS) and may further not be representative to best fit the test set.$\blacktriangleleft$}

%

\section{Discussions}
\label{sec:dis}
We enumerate a few insights from our experiments with representation learning models and discuss some threats to validity.

\subsection{Experimental insights}
[{\em Code-oriented embedding models may not yield the best embeddings for training predictors.}] Our experiments have revealed that the BERT model which was pre-trained on Wikipedia is yielding the best recall in the identification of incorrect patches. 
There are several possible reasons to that: Bert implements the deepest neural network and builds on the largest training data. Its performance suggests that code-oriented embeddings should aim for being accurate with small training datasets in order to become competitive against BERT. While we were completing the experiments, a pre-trained CodeBERT~\cite{feng2020codebert} model has been released (on April 27). In future work, we will investigate its relevance for producing embeddings that may yield higher performance in patch correctness prediction. In any case, we note that CC2Vec provided the best embeddings for yielding the best recall in identifying correct. patches (using similarity thresholds). This suggests that future research should investigate the value of merging different representations or combining the eventual prediction probabilities to improve performance on identifying correct patches and excluding incorrect patches.

\begin{figure}[!h]
\vspace{-1mm}
	\includegraphics[width=1\linewidth]{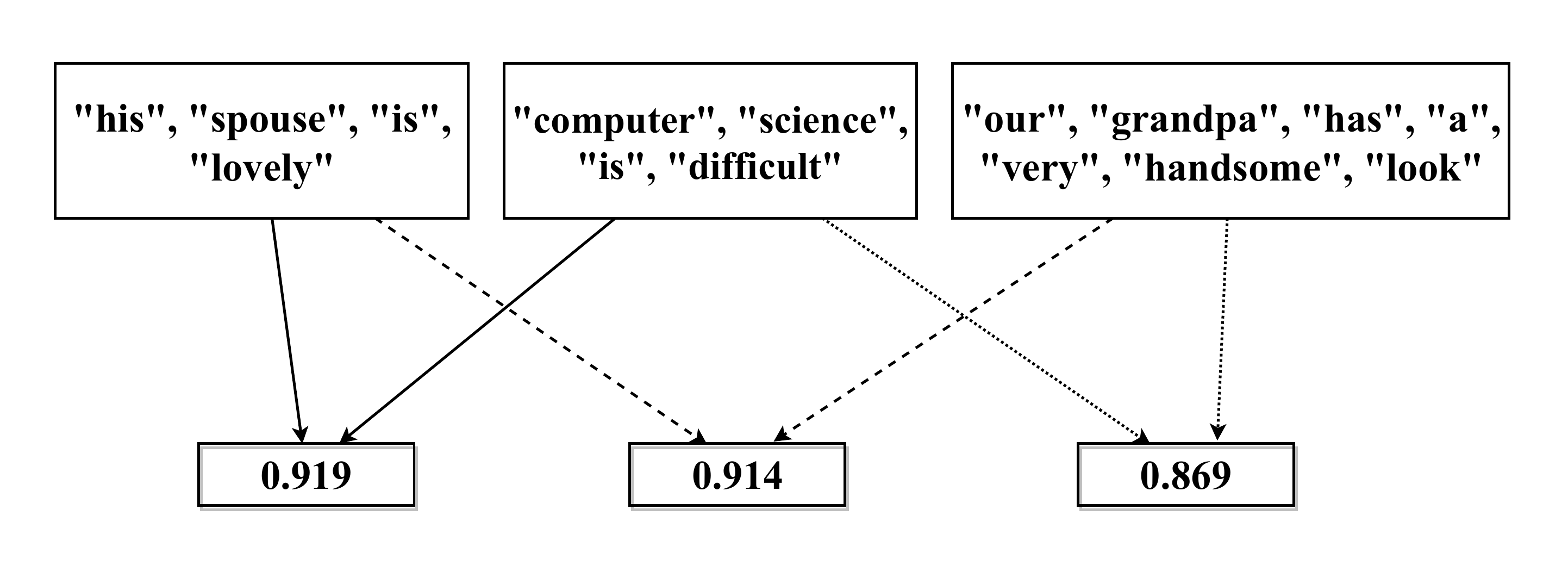}
	\caption{Close cosine similarity scores with small-sized inputs for BERT embedding model.}
	\label{fig:input-size}
\end{figure}

\vspace{0.1cm}
\noindent
[{\em The small sizes of the code fragments lead to similar embeddings.}].
Figure~\ref{fig:input-size} illustrates the different cosine similarity scores that can be obtained for the BERT embeddings of different pairs of short sentences. Although the sentences are semantically (dis)similar, the cosine similarity scores are quite close. This explains why recalling correct patches based on a similarity threshold was a failed attempt ($\sim 5\%$ for APR-generated patches for. Defects4J+Bears+Bugs.jar bugs). Nevertheless, experimental results demonstrated that deep learned features were relevant for learning to discriminate.

\noindent
[{\em Embeddings are most suitable when applied to simple ML algorithms.}]
Because embeddings are yielded from neural networks, they are actually formed by complex crossed features. When they are fed to a complex discriminant model such as Random Forrest, it may lead to overfitting with small datasets. Our experiments however show that simple Logistic Regression yields the best AUC,  suggesting that this learner was able to better identifying discriminating features for the prediction task.

\vspace{-1mm}
\subsection{Threats to validity}

Our empirical study carries a number of threats to validity that we have tried to mitigate.

\vspace{0.1cm}
\noindent
{\sc \em Threats to External Validity.} There are a variety of representation learning models in the literature. A threat to validity of our study is that we may have a selection bias by considering only four embedding models. We have mitigated this threat by considering representative models in different scenarios (pre-trained vs retrained, code change specific vs natural language oriented).

Another threat to validity is related to the use of Defects4J data in evaluating the ML classifiers. This choice however was dictated by the data available and the aim to compare against related work.

Finally, with respect to the explored models, the attention system of CC2Vec requires some execution parameters to perform well.
Since the relevant code was not available, we use use a non-attention version instead, potentially making CC2Vec embeddings be under-performing. We release the artifacts for future comparisons by the research community. 

\vspace{0.1cm}
\noindent
{\sc \em Threats to Internal Validity.}
A major threat to internal validity lies in the manual assessment heuristics that we applied to the RepairThemAll-generated dataset. We may have misclassified some patches due to mistakes or conservatism. 
This threat however holds for all APR work that relies on manual assessment. 
We mitigate this threat by following clear and reproducible decision criteria, and by releasing our labelled datasets for the community to review\footnote{see: \url{https://github.com/SerVal-DTF/DL4PatchCorrectness}}.

\vspace{0.1cm}
\noindent
{\sc \em Threats to Construct Validity.}
For our experiment, the considered embedding models are not perfect and they may have been under-trained for the prediction task that we envisioned. For this reason, the results that we have reported are likely
an under-estimation of the capability of representation learning models to capture discriminative features for the prediction of patch correctness.
 Our future studies on representation learning will
address this threat by considering different re-training experiments.
\section{Related Work}
\label{sec:relatedWork}

\paragraph{\bf Analyzing Patch Correctness:}
To assess the performance of fixing bugs of repair tools and approaches, 
checking the correctness of patches is key, but not trivial.
However, this task was largely ignored or unconcerned in the community until the analysis study of patch correctness conducted by Qi~{\em et~al.}~\cite{qi2015analysis}.
Thanks to their systematic analysis of the patches reported by three generate-and-validate program repair systems (i.e., GenProg, RSRepair and AE), they shown that the overwhelming majority of the generated patches are not correct but just overfit the test inputs in the test suites of buggy programs.
In another study, Smith~{\em et~al.}~\cite{smith2015cure} uncover that patches generated with lower coverage test suites overfit more.
Actually, these overfitting patches often simply break under-tested functionalities, and some of them even make the ``patched'' program worse than the un-patched program. 
Since then, the overfitting issue has been widely studied in the literature.
For example, Le~{\em et~al.}~\cite{le2018overfitting} revisit the overfitting problem in semantics-based APR systems.
In~\cite{le2019reliability}, they further assess the reliability of authors and automated annotations in assessing patch correctness. They recommend to make  publicly available to the community the  patch correctness evaluations of the  authors. 
Yang and Yang~\cite{yang2020exploring} explore the difference between the runtime behavior of programs patched with developer's patches and those by APR-generated plausible patches. 
They unveil that most APR-generated plausible patches lead to different runtime behaviors compared to correct patches.

\paragraph{\bf Predicting Patch Correctness:}
To predict the correctness of patches, one of the first explored research directions relied on the idea of augmenting test inputs, i.e., more tests need to be proposed. 
Yang~{\em et~al.}~\cite{yang2017better} design a framework to detect overfitting patches. 
This framework leverages fuzz strategies on existing test cases in order to automatically generate new test inputs. In addition, it leverages additional oracles (i.e., memory-safety oracles) to improve the validation of APR-generated patches.
In a contemporary study, Xin and Reiss~\cite{xin2017identifying} also explored to generate new test inputs, with the syntactic differences between the buggy code and its patched code, for validating the correctness of APR-generated patches.
As complemented by Xiong~{\em et~al.}~\cite{xiong2018identifying}, they proposed to assess the patch correctness of APR systems by leveraging the automated generation of new test cases and measuring behavior similarity of the failing tests on buggy and patched programs.

Through an empirical investigation, Yu~{\em et~al.}~\cite{yu2019alleviating} summarized two common overfitting issues: incomplete fixing and regression introduction.
To assist alleviating the overfitting issue for synthesis-based APR systems,
they further proposed \texttt{UnsatGuided} that relies on additional generated test cases to strengthen patch synthesis, and thus reduce the generation of incorrect overfitting patches.

Predicting patch correctness with thanks to an augmented set of test cases heavily relies on the quality of tests. In practice, tests with high coverage might be unavailable~\cite{ye2019automated}.
In our paper, we do not rely on any new test cases to assess patch correctness, but leverage  representation learning techniques to build representation vectors for buggy and patched code of APR-generated patches.

To predict overfitting patches yielded by APR tools, Ye~{\em et~al.}~\cite{ye2019automated} propose ODS, an overfitting detection system. 
ODS first statically extracts 4,199 code features at the AST level from the buggy code and generated patch code of APR-generated patches. 
Those features are fed into 
three machine learning algorithms (logistic regression, KNN, and random forest) 
to learn an ensemble probabilistic model for classifying and ranking potentially overfitting patches. 
To evaluate the performance of ODS, the authors considered 
19,253 training samples and 713 testing samples from the  Durieux~{\em et~al.} empirical study~\cite{durieux2019empirical}. With these settings,
ODS is capable of detecting 57\% of overfitting patches.
The ODS approach relates to our study since both leverage machine learning and static features. However, ODS only relies on manually identified features which may not generalize to other programming languages or even other datasets. 

In a  recent work, Csuvik~{\em et~al.}~\cite{csuvik2020utilizing} exploit the textual and structural similarity between the buggy code and the APR-patched code with two representation learning models (BERT~\cite{devlin2019bert} and Doc2Vec~\cite{le2014distributed}) by considering three patch code representation (i.e., source code, abstract syntax tree and identifiers).
Their results show that the source code representation is likely to be more effective in correct patch identification than the other two representations, 
and the similarity-based patch validation can filter out incorrect patches for APR tools.
However, to assess the performance of the approach, only 64 patches from QuixBugs~\cite{ye2019comprehensive} have been considered (including 14 in-the-lab bugs). This low number of considered patches raises questions about the generalization of the approach for fixing bugs in the wild. 
Moreover, unlike our study, new representation learning models (code2vec~\cite{alon2019code2vec} and CC2Vec~\cite{hoang2020cc2vec}) dedicated to code representation have not been exploited.

\paragraph{\bf Representation Learning for Program Repair Tasks:}
In the literature, representation learning techniques have been widely explored to boost program repair tasks.
Long and Rinard explored the topic of learning correct code for patch generation~\cite{long2016automatic}. 
Their approach learns code transformation for three kinds of bugs from their related human-written patches.
After mining the most recent 100 bug-fixing commits from each of the 500 most popular Java projects, Soto and Le Goues~\cite{soto2018using} have built a probabilistic model to predict bug fixes for program repair.
To identify stable Linux patches, Hoang~{\em et~al.}~\cite{hoang2019patchnet} proposed a hierarchical deep learning-based method with features extracted from both commit messages and commit code. 
Liu~{\em et~al.}~\cite{liu2018mining2} and Bader~{\em et~al.}~\cite{bader2019getafix} proposed to learn recurring fix patterns from human-written patches and suggest fixes. 
Our paper is not aiming at proposing a new automated patch generation approach. We indeed  rather focus on assessing representation learning techniques for predicting correctness of patches generated by program repair tools.

\section{Conclusion}
\label{sec:conc}
In this paper, we investigated the feasibility of statically predicting patch correctness by leveraging representation learning models and supervised learning algorithms. The objective is to provide insights for the APR research community towards improving the quality of repair candidates generated by APR tools.
To that end, we, first investigated the use of different distributed representation learning to capture the similarity/dissimilarity between buggy and patched code fragments.
These experiments gave similarity scores that substantially differ for across embedding models such as BERT, Doc2Vec, code2vec and CC2Vec.
Building on these results and in order to guide the exploitation of code embeddings in program repair pipelines, we investigated in subsequent experiments the selection of cut-off similarity scores to decide which APR-generated patches are likely incorrect.
This allowed us to filter out between 31.5\% and 94.9\% incorrect patches based on brute cosine similarity scores.
Finally, we investigated the discriminative power of the deep learned features by training machine learning classifiers to predict correct Patches. DecisionTree, Logistic Regression and Naive Bayes are tried with code embeddings from BERT, Doc2Vec and CC2Vec.
Logistic Regression with BERT embeddings yielded very promising performance on patch correctness
prediction with metrics like F-Measure at 0.72\% and AUC at 0.8\% on a labeled deduplicated dataset of 1000 patches. We further showed that the performance of these models on static features is promising when comparing against the state of the art (PATCH-SIM~\cite{xiong2018identifying}), which uses dynamic execution traces. Experimental results suggests that the deep learned features can be complementary to hand-crafted features (such as those engineered by ODS~\cite{ye2019automated}).

\vspace{0.2cm}
\noindent
{\bf Availability.} All artifacts of this study are available in the following public repository:
\begin{center}
	\url{https://github.com/SerVal-DTF/DL4PatchCorrectness}
\end{center}

\section*{Acknowledgements}{
This work is supported by the Project 1015-YAH20102, the National Natural Science Foundation of China (Grant No.61802180), the Natural Science Foundation of Jiangsu Province (Grant No.BK20180421), the National Cryptography Development Fund (Grant No.MMJJ20180105) and the Fundamental Research Funds for the Central Universities (Grant No.NE2018106).}

\balance
\bibliographystyle{ACM-Reference-Format}
\bibliography{bib/references}

\end{document}